\documentclass[twocolumn]{aastex63}
\usepackage{xcolor}
\begin{document}

\title{The Rate, Amplitude and Duration of Outbursts from Class 0 Protostars in  Orion}
\correspondingauthor{S. T. Megeath}
\email{s.megeath@utoledo.edu}
\author{Wafa Zakri}
\affiliation{Ritter Astrophysical Research Center, Dept. of Physics and Astronomy, University of Toledo, Toledo, OH 43606, USA}
\affiliation{ Department of Physics, Jazan University, Jazan, Saudi Arabia}
\author[0000-0001-7629-3573]{S. T. Megeath}
\affiliation{Ritter Astrophysical Research Center, Dept. of Physics and Astronomy, University of Toledo, Toledo, OH 43606, USA}
\author[0000-0002-3747-2496]{William J. Fischer}
\affiliation{Space Telescope Science Institute, 3700 San Martin Drive, Baltimore, MD 21218, USA}
\author[0000-0002-6447-899X]{Robert Gutermuth}
\affiliation{Department of Astronomy, University of Massachusetts, Amherst, MA 01003  USA}
\author[0000-0001-9800-6248]{Elise Furlan}
\affiliation{NASA Exoplanet Science Institute, Caltech/IPAC, Mail Code 100-22, 1200 E. California Blvd., Pasadena, CA 91125, USA}
\author[0000-0003-1430-8519]{Lee Hartmann}
\affiliation{Department of Astronomy, University of Michigan, 1085 S. University Street, Ann Arbor, MI 48109, USA}
\author[0000-0003-3682-854X]{Nicole Karnath}
\affiliation{SOFIA-USRA, NASA Ames Research Center, MS 232-12, Moffett Field, CA 94035, USA}
\author[0000-0002-6737-5267]{Mayra Osorio}
\affiliation{Instituto de Astrofisica de Andalucia, CSIC, Camino Bajo de Hu\'etor 50, E-18008 Granada, Spain}
\author[0000-0002-6872-2582]{Emily Safron}
\affiliation{Department of Physics and Astronomy, Lousiana State University, Baton Rouge, LA 70803 USA}
\author[0000-0002-5812-9232]{Thomas Stanke}
\affiliation{European Southern Observatory, Karl-Schwarzschild-Str. 2, D-85748 Garching b. M\"unchen, Germany}
\author[0000-0003-2300-8200]{Amelia M.\ Stutz}
\affiliation{Departamento de Astronom\'{i}a, Universidad de Concepci\'{o}n,Casilla 160-C, Concepci\'{o}n, Chile}
\affiliation{Max-Planck-Institute for Astronomy, K\"{o}nigstuhl 17, 69117 Heidelberg, Germany}
\author[0000-0002-6195-0152]{John J. Tobin}
\affiliation{National Radio Astronomy Observatory, 520 Edgemont Rd., Charlottesville, VA 22903, USA}
\author{Thomas S. Allen}
\affiliation{Lowell Observatory, 1400 W. Mars Hill Road, Flagstaff AZ 86001, USA}
\author[0000-0002-6136-5578]{Sam Federman}
\affiliation{Ritter Astrophysical Research Center, Dept. of Physics and Astronomy, University of Toledo, Toledo, OH 43606, USA}
\author[0000-0002-2667-1676]{Nolan Habel}
\affiliation{Ritter Astrophysical Research Center, Dept. of Physics and Astronomy, University of Toledo, Toledo, OH 43606, USA}
\author[0000-0002-3530-304X]{P. Manoj}
\affiliation{Department of Astronomy and Astrophysics, Tata Institute of Fundamental Research, Homi Bhabha Rd, Colaba, Mumbai 400005, India}
\author[0000-0002-0554-1151]{Mayank Narang}
\affiliation{Department of Astronomy and Astrophysics, Tata Institute of Fundamental Research, Homi Bhabha Rd, Colaba, Mumbai 400005, India}
\author[0000-0002-0557-7349]{Riwaj Pokhrel}
\affiliation{Ritter Astrophysical Research Center, Dept. of Physics and Astronomy, University of Toledo, Toledo, OH 43606, USA}
\author[0000-0001-6381-515X]{Luisa Rebull}
\affiliation{Infrared Science Archive (IRSA), IPAC, 1200 E. California Blvd., California Institute of Technology, Pasadena, CA 91125, USA}
\author[0000-0002-9209-8708]{Patrick D. Sheehan}
\affiliation{Center for Interdisciplinary Exploration and Research in Astronomy, 1800 Sherman Rd., Evanston, IL 60202, USA}
\author[0000-0001-8302-0530]{Dan M. Watson}
\affiliation{Department of Physics and Astronomy, University of Rochester, Rochester, NY 14627, USA}

\vskip 0.1 in 
\begin{abstract}
At least half of a protostar's mass is accreted in the Class 0 phase, when the central protostar is deeply embedded in a dense, infalling envelope. We present the first systematic search for outbursts from Class 0 protostars in the Orion clouds. Using photometry from  Spitzer/IRAC spanning 2004 to 2017, we detect three outbursts from Class 0 protostars with $\ge 2$ mag changes at 3.6 or 4.5 $\mu$m. This is comparable to the magnitude change of a known protostellar FU Ori outburst. Two are newly detected bursts from the protostars HOPS 12 and 124. The number of detections implies that Class 0 protostars burst every 438 yr,  with a 95\% confidence interval of 161 to 1884 yr. Combining Spitzer and WISE/NEOWISE data spanning 2004-2019, we show that the bursts persist for more than nine years with significant variability during each burst.  Finally, we use $19-100$ $\mu$m photometry from SOFIA, Spitzer and Herschel to measure the amplitudes of the bursts. Based on the burst interval, a duration of 15 yr, and the range of observed amplitudes, 3--100\% of the mass accretion during the Class 0 phase occurs during bursts.  In total, we show that bursts from Class 0 protostars are as frequent, or even more frequent, than those from  more evolved protostars. This is consistent with bursts being driven by instabilities in disks triggered by rapid mass infall. Furthermore, we find that bursts may be a significant, if not dominant, mode of mass accretion during the Class 0 phase.
\end{abstract}



\section{INTRODUCTION}

 The outburst of HOPS~383, identified in multi-epoch Spitzer and WISE observations, was the first detection of a luminous outburst toward a Class 0 protostar \citep{2015ApJ...800L...5S}. This outburst appears similar to the outbursts driven by rapid increases in mass accretion detected toward more evolved protostars. Class 0 protostars sample the earliest phase of protostellar evolution \citep{1993ApJ...406..122A, 2016ApJS..224....5F}, and more than half of mass accretion is thought to occur in the Class 0 phase \citep{2017ApJ...840...69F}.  If outbursts are common for Class 0 protostars, they may contribute a substantial fraction of a star's mass. Indeed, simulations of protostars and modeling of chemical abundances in protostellar envelopes predict that outbursts are frequent in the Class 0 phase \citep{2014ApJ...795...61B,2015A&A...577A.102V,2019ApJ...884..149H}. 

The luminosity functions of protostars also suggest bursts are a common, if not dominant, mode of accretion for protostars. In particular, models of protostellar evolution invoking isothermal sphere collapse require luminosities higher than those observed.  This discrepancy, known as the protostellar luminosity problem, can be resolved if most of the stellar mass is accreted during bursts \citep{1994AJ....108..251K,2012ApJ...747...52D}. There are alternative models of infall and accretion, however, that do not require episodic bursts to reproduce the luminosity function \citep{2011ApJ...736...53O}. 

Establishing the importance of bursts for mass accretion requires multi-epoch, photometric surveys that provide direct measurements of their rate, amplitudes, and durations. While ground-based 2~\micron\ observations can detect bursts from Class I and flat-spectrum protostars \citep[e.g.][]{2017MNRAS.465.3011C}, reliable detections of Class~0 protostars and their bursts requires observations at $> 2.5$~\micron.  The Spitzer Space Telescope \citep{2004ApJS..154....1W} detected all but the youngest, most deeply embedded protostars \citep{2012AJ....144...31K,2013ApJ...767...36S}.
The Spitzer survey of the Orion molecular clouds in 2004-2005 provided baseline measurements of protostellar flux densities from 3.6 to 70~$\mu$m \citep{2012AJ....144..192M}. An extension of the Spitzer survey in 2008, still during the cryogenic mission, led to detections of outbursts in the overlapping regions between the 2008 and 2004/2005 MIPS 24~$\mu$m maps. This revealed outbursts of the flat-spectrum protostar V2775 Ori \citep[HOPS 223;][]{2011A&A...526L...1C,2012ApJ...756...99F} and the Class~0 protostar HOPS 383 \citep{2015ApJ...800L...5S}. Examination of the WISE survey 3.4--22~$\mu$m photometry in 2010 showed no additional bursts, leading to an estimated interval between bursts of 1000 yr for each protostar \citep{2019ApJ...872..183F}.

In this paper, we search for outbursts from the 92 identified Class 0 protostars in the Orion clouds by combining the 2004 cryogenic data of Orion with warm mission data taken during the YSOVAR exploration program and the OrionTFE (The Final Epoch) survey, which duplicated the original survey fields in 2016--2017. These data provide sparsely sampled 3.6 and 4.5~$\mu$m light curves that span 13 yr with a higher angular resolution and sensitivity than WISE/NEOWISE. 

We announce the discovery of outbursts from the Class~0 protostars HOPS 12 and  124, and show the light curve covering the entire burst for the previously known Class 0 protostar HOPS 383.  We supplement the light curves with additional Spitzer and WISE/NEOWISE photometry. We use 19--100~$\mu$m mid to far-IR data to estimate the change in luminosity during each burst, with SOFIA providing  additional epochs to earlier Spitzer and Herschel observations.

\section{Observations and Data Analysis}

\subsection{Spitzer Cryogenic and Warm Mission Data}
\label{sec:spitzer_data}

\begin{deluxetable}{ccc}
\tablecaption{Spitzer/IRAC Photometry\label{t.irac_phot}}
\tablecolumns{3}
\tablehead{\colhead{} & \multicolumn{2}{c}{Flux Density (mJy)} \\[-0.2cm] \colhead{Date} & \colhead{3.6 $\micron$} & \colhead{4.5 $\micron$}}
\startdata
\cutinhead{HOPS 12}
2004 Mar  9 & $0.45\pm0.02$ & $2.63\pm0.04$ \\
2004 Oct 12 & $0.44\pm0.02$ & $2.30\pm0.04$ \\
2009 Nov 11$^1$ & $1.65\pm0.04^2$ & $14.2\pm0.4^2$ \\
2010 Nov 19$^3$ & $1.85\pm0.05^2$ & $14.2\pm0.7^2$ \\
2017 Jan 11 & $1.71 \pm 0.03 $ & $15.6 \pm 0.1$ \\
2017 Jun  1 & $1.82 \pm 0.03$ & $16.2 \pm 0.1$ \\
2019 Jul 21 & $0.92 \pm 0.02 $& $9.8 \pm 0.1$ \\
\cutinhead{HOPS 124}
2004 Feb 17 & $3.59 \pm 0.11$ & $16.9 \pm 0.3$ \\
2004 Feb 18 & $3.64 \pm 0.11$ & $17.4 \pm 0.2$ \\
2004 Oct  8 & $2.65 \pm 0.13$ & $13.6 \pm 0.4$ \\
2004 Oct 27 & $2.69 \pm 0.11$ & $14.5 \pm 0.3$ \\ 
2017 Jan 1 & $18.7 \pm 0.3$ & $122 \pm 2$ \\
2017 Jan 6 & $22.0 \pm 0.3$ & $129 \pm 2$ \\
2017 Jun  1 & $18.3 \pm 0.3$ & $112 \pm 2$ \\
2017 Jun  6 & $19.8 \pm 0.4$ & $126 \pm 1$ \\
2019 Jul 15 & $13.4 \pm 0.2$ & $84.0 \pm 0.9$ \\
\cutinhead{HOPS 383}
2004 Mar  9 & $\cdots$ & $0.30 \pm 0.02$ \\
2004 Oct 12 & $\cdots$ & $1.34 \pm 0.03$ \\
2009 Nov 12$^1$ & $\cdots$ & $6.17 \pm 0.07^2$ \\
2010 Nov 19$^3$ & $\cdots$ & $6.46 \pm 0.20^2$ \\
2017 Jan 11 & $\cdots$ & $1.23 \pm 0.03$ \\
2017 Jun  1 & $\cdots$ & $0.57 \pm 0.03$ \\
2019 Jul 15 & $\cdots$ & $0.32 \pm 0.02$ \\
2019 Jul 21 & $\cdots$ & $0.33 \pm 0.02$ \\
\enddata
\tablenotetext{1}{Light curve of first YSOVAR campaign taken from 2009 Oct 23 to Dec 01}
\tablenotetext{2}{Median value of light curve $\pm$ median absolute deviation of light curve}
\tablenotetext{3}{Light curve of second YSOVAR  campaign taken from 2010 Oct 29 to Dec 09}
\end{deluxetable}

Spitzer initially surveyed Orion with the IRAC camera \citep{2004ApJS..154...10F} during the cryogenic (hereafter:\ cryo) mission, observing most protostars in two epochs separated by six months in 2004, with single-epoch field extensions in 2007 and 2009 \citep{2012AJ....144..192M,2012AJ....144...31K}. With the onset of the warm mission, the YSOVAR exploration program observed the Orion Nebula Cluster (ONC) during 80 epochs in 2009 and 2010 \citep{2011ApJ...733...50M,2014AJ....148...92R}. The Orion:\ The Final Epoch (hereafter:\ OrionTFE) program then duplicated the cryo field in two epochs in 2016 and 2017. Finally, during the Spitzer Beyond phase of the mission, Spitzer observed YSOs in 2019 that had been found to vary by $\ge 1$~mag between the cryogenic and OrionTFE programs.

We extracted photometry from images downloaded from the Spitzer Heritage Archive and processed through a custom IDL program developed for cryo mission data and then later adapted for the warm mission \citep{2009ApJS..184...18G,2011ApJ...733...50M}.  At the positions of YSOs taken from  \citet{2012AJ....144..192M,2016AJ....151....5M}, we performed aperture photometry on the individual frames using the ``aper'' utility from the IDL Astronomy Users' Library \citep{1993ASPC...52..246L}. The aperture radius was $2.4''$ and the sky annulus extended $2.4''$ to $7.2''$. The adopted zero points in DN s$^{-1}$ at the IRAC native pixel scale were 19.455 and 18.699 for the cryo data and 19.306 and 18.669 for the warm data in the 3.6 and 4.5~$\mu$m bands, respectively \citep{2008ApJ...674..336G,2011ApJ...733...50M}. Magnitudes were converted to flux densities using the zero magnitude fluxes from \citet{2005PASP..117..978R}. The Spitzer/IRAC photometry is shown in Table~\ref{t.irac_phot}. We present only median values of the YSOVAR photometry for each of the two Orion campaigns; the light curves are available in \citet{2011ApJ...733...50M}.

For the comparison of the HOPS~12 Spitzer photometry to the WISE/NEOWISE data, we degraded the Spitzer data from the cryo and warm mission to a resolution similar to that of WISE. To do this, the Spitzer 3.6 and 4.5~$\mu$m mosaics were convolved with a kernel  to reproduce the larger WISE 3.4 and 4.6~$\mu$m PSFs. We then extracted photometry from the Spitzer images with an aperture radius of 12.4\arcsec\ and a sky annulus extending from 12.4\arcsec\ to 15.1\arcsec. The data were calibrated by comparing the photometry of isolated stars in the convolved images with the photometry of \citet{2012AJ....144..192M}. 

\subsection{WISE and NEOWISE Data}

We used WISE mission 3.4 and 4.6 $\mu$m data to supplement the Spitzer photometry \citep{2010AJ....140.1868W, 2011ApJ...731...53M}. For HOPS~12, the photometry was obtained from the WISE All-Sky Database, the WISE 3-Band Cryo Database, and the NEOWISE-R single-image photometry data. For the single-image photometry, we averaged all points taken within a single week. To separate HOPS~124 from the neighboring HOPS~125, we performed PSF fitting on unWISE images for every individual epoch of the WISE and NEOWISE mission from 2010 to 2017 \citep{2014AJ....147..108L,2017AJ....154..161M}. Using the DAOPHOT implementation in the IDL Astronomy Users' Library, we simultaneously fit the PSF of HOPS~124 and the overlapping PSFs of nearby point sources \citep{1993ASPC...52..246L}. 

\subsection{SOFIA FORCAST Photometry}

HOPS 12, HOPS 124, HOPS 383 and their surroundings were observed with FORCAST aboard SOFIA in programs 04\_0134 and 07\_0200. HOPS 12 was observed at 25.3 and 37.1 \micron\ on 2019 October 16 and at 19.7, 25.3, 31.5, and 37.1 \micron\ on 2019 October 25. HOPS 124 was observed at the same four wavelengths on 2019 October 16. HOPS 383 was observed at 25.3 and 37.1 \micron\ on 2016 February 11. The observations were performed with two-position chopping and nodding 180$^\circ$ from the chop direction (C2N/NMC mode). 

We used the flux-calibrated (Level 3) images provided by the SOFIA Science Center, with 0.768\arcsec\ pixels. We obtained flux densities of the targets in 8 or 10 pixel (6.14\arcsec\ or 7.68\arcsec) apertures with subtraction of the modal sky signal in an annulus extending from 10 pixels to 12 pixels (7.68\arcsec\ to 9.22\arcsec); these were determined from inspection of the point-spread functions. The color corrections needed to compare the FORCAST and MIPS 24~$\mu$m photometry are described in Appendix~A. The extraction of the MIPS 24~$\mu$m fluxes is described in \citet{2012AJ....144..192M}.

\subsection{Far-IR Data from Herschel, Spitzer and SOFIA}

Data from PACS on Herschel \citep{2010A&A...518L...2P,2010A&A...518L...1P}, MIPS on Spitzer \citep{2004ApJS..154...25R}, and HAWC+ on SOFIA \citep{2018JAI.....740008H} measured the 70--100~$\mu$m far-IR fluxes of the three protostars before, during, and/or after the bursts.  \citet{2020ApJ...905..119F} describe the acquisition, data processing, and photometric measurement of the PACS 70 \micron\ images, which come from the Herschel Orion Protostar Survey \citep[HOPS;][]{2016ApJS..224....5F}. The 100 \micron\ PACS photometry used for HOPS 383 is from Herschel Gould Belt Survey data \citep{2010A&A...518L.102A} and was extracted using the approach described in \citet{2013ApJ...767...36S}. For HOPS~12, we separated the 70~$\mu$m PACS fluxes of the blended eastern and western sources using the IDL implementation of DAOPHOT \citep{1993ASPC...52..246L}. The positions and fluxes of the sources were then tweaked to minimize the artifacts in the subtracted images as judged by eye. HOPS~12 was also observed pre-outburst by MIPS at 70~$\mu$m; photometry for this source was obtained using the procedure of \citet{2012ApJ...756...99F}, which uses 70 \micron\ PACS data to calibrate 70 \micron\ MIPS data.

HOPS 383 was imaged at 89~\micron\ with 8\arcsec\ resolution by HAWC+ aboard SOFIA in program 07\_0200. Observations were obtained in total-intensity scan-mapping mode on 2019 September 10. We extracted photometry from the flux-calibrated (Level 3) image provided by the SOFIA Science Center. We used an 11 pixel (17.05\arcsec) aperture and subtracted the modal sky signal in an annulus extending from 11 pixels to 16 pixels (17.05\arcsec\ to 24.80\arcsec); these were determined from inspection of the point-spread function of the brighter HOPS 90.

\begin{figure*}[t!]
\centering
\includegraphics[height=9.5cm, width=18  cm]{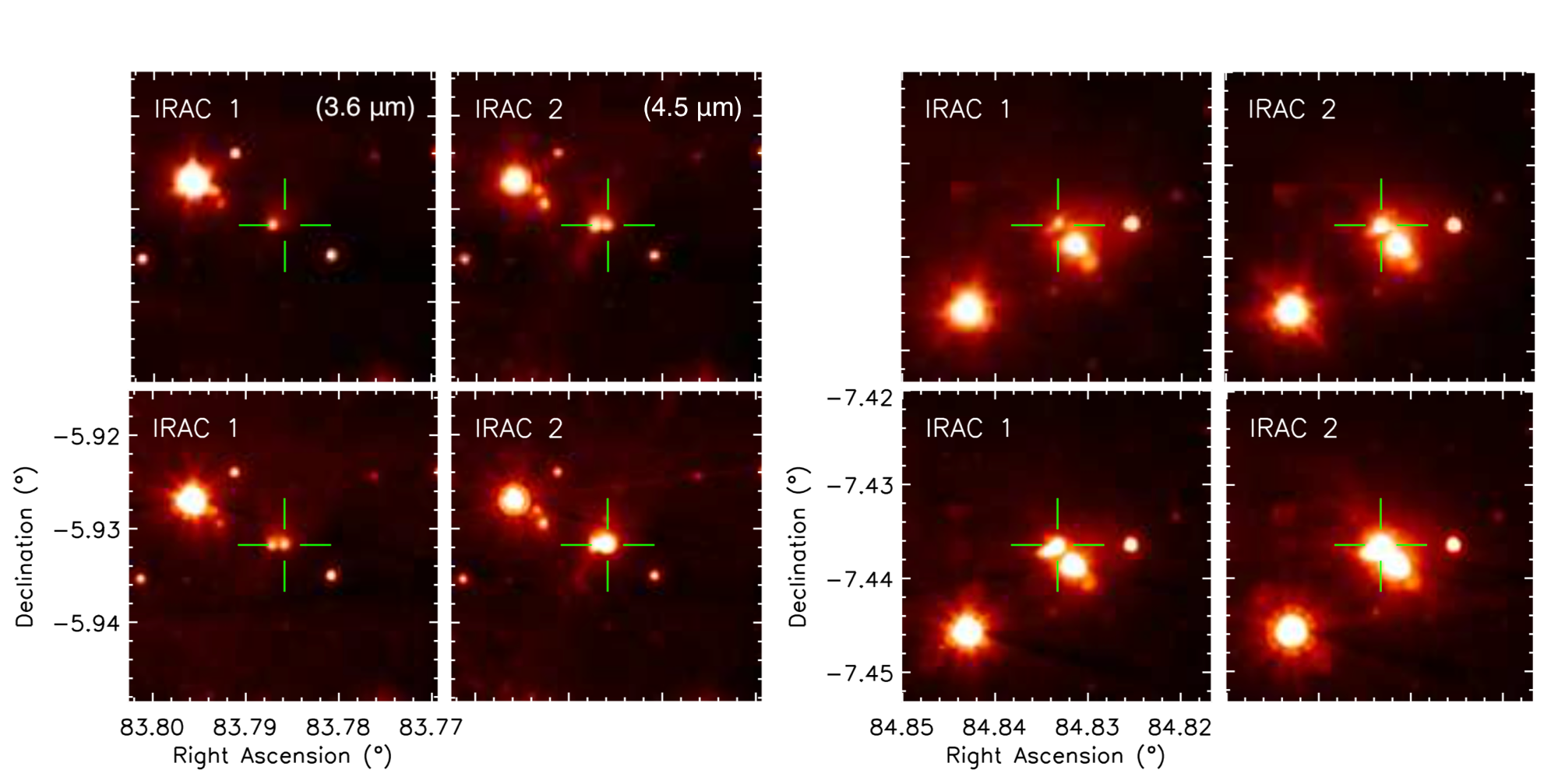}
\caption{Spitzer 3.6 and 4.5~$\mu$m images of the HOPS~12 outburst (left, $\alpha=83.78584^\circ$, $\delta=-5.93174^\circ$ J2000) and the HOPS~124 outburst (right, $\alpha=84.83327^\circ$, $\delta=-7.43644^\circ$ J2000) with the locations of the protostars marked. The 2004 cyro images (top) show the protostars pre-outburst, while the OrionTFE images (bottom) show the protostars during the bursts.}
\label{fig:image}
\end{figure*}

The 19.7 to 100 \micron\ fluxes from Spitzer, SOFIA, and Herschel are shown in Table~\ref{t.long_phot}.

\begin{deluxetable}{cCcc}
\tablecaption{Photometry from 19.7 to 100 \micron\label{t.long_phot}}
\tablecolumns{4}
\tablehead{\colhead{Wavelength} & \colhead{Flux Density} & \colhead{Instrument} & \colhead{Date} \\[-0.2cm] \colhead{(\micron)} & \colhead{(mJy)} & \colhead{} & \colhead{}}
\startdata
\cutinhead{HOPS 12}
19.7                & 276\pm22      & FORCAST & 2019 Oct 25 \\
24                  & 276\pm14      & MIPS    & 2004 Mar 20 \\
25.3                & 1130\pm60     & FORCAST & 2019 Oct 16 \\
25.3                & 1140\pm90     & FORCAST & 2019 Oct 25 \\
31.5                & 2560\pm30     & FORCAST & 2019 Oct 25 \\
37.1                & 3160\pm140    & FORCAST & 2019 Oct 16 \\
37.1                & 3170\pm200    & FORCAST & 2019 Oct 25 \\
70                  & 5780\pm580          & MIPS    & 2004 Mar 20 \\
70                  & 15600\pm800   & PACS    & 2010 Sep  09 \\
\cutinhead{HOPS 124}
19.7                & 3170\pm60     & FORCAST & 2019 Oct 16 \\
24                  & 1710\pm90     & MIPS    & 2005 Apr 02 \\
25.3                & 15600\pm200   & FORCAST & 2019 Oct 16 \\
31.5                & 38100\pm100   & FORCAST & 2019 Oct 16 \\ 
37.1                & 57800\pm400   & FORCAST & 2019 Oct 16 \\
\cutinhead{HOPS 383}
24                  & 5.08\pm0.27   & MIPS    & 2004 Mar 20 \\
24                  & 178\pm9       & MIPS    & 2008 Apr 19 \\
25.3                & 279\pm40      & FORCAST & 2016 Feb 11 \\
37.1                & 569\pm79      & FORCAST & 2016 Feb 11 \\
70                  & 13100\pm200   & PACS    & 2010 Sep 10 \\
89                  & 2160\pm170    & HAWC$+$ & 2019 Sep 10 \\
100                 & 25400\pm1400  & PACS    & 2010 Oct 08  \\
\enddata
\end{deluxetable}
\vskip -0.5 in

\section{RESULTS}

In a comparison of the cryo, OrionTFE, and YSOVAR data for 319 protostars, we identified five protostars with bursts of $\ge 2$~mag in the 3.6 or 4.5 $\micron$ Spitzer bands. This includes the previously known outbursts from the flat-spectrum (hereafter:\ FS) protostar V2775~Ori, or HOPS~223 \citep{2011A&A...526L...1C,2012ApJ...756...99F}, the Class I protostar HOPS~41 \citep[][W.~Zakri et al., in prep.]{2021ApJ...920..132P}, and the Class 0 protostar HOPS~383 \citep{2015ApJ...800L...5S}.  The remaining  objects are newly identified outbursts of the Class 0 protostars HOPS~12 and HOPS~124. The focus of this paper is the Class 0 outbursts.

The 1-2.5~$\mu$m spectrum of V2775~Ori shows it to be one of the lowest amplitude FU Ori outbursts known \citep{2012ApJ...756...99F}. We thus adopt the 2~mag threshold to focus on bursts with amplitudes comparable to known FU Ori outbursts in more evolved protostars. Furthermore, in the Orion sample, only bursts show variations at  3.6 or 4.5 $\micron$ that exceed 2~mag. In a companion paper, we will show that protostars with smaller amplitude variations exhibit a more diverse range of behaviors, including bursts, fluctuations and fades \citep[W.~Zakri et al., in prep.; see also][]{2021ApJ...920..132P}.   

The outbursts of HOPS~12 and 124 are shown by the cryo and OrionTFE images in Figure~\ref{fig:image}.  HOPS~12 is found in the ONC and was consequently observed in two month-long campaigns by the YSOVAR program \citep[see][]{2014AJ....148...92R}. HOPS~124 is located in the Lynds~1641 cloud \citep{2008hsf1.book..621A}. Both protostars have nearby YSOs that may be companions.  HOPS~12 is separated by 6.46'' (projected separation of 2,510~au at 389~pc) from a YSO later resolved into a close binary by ALMA and VLA data \citep{2020ApJ...890..130T}. We refer to the outbursting source as HOPS~12~West and the close binary as HOPS~12~East. HOPS~124 is 9.47'' \citep[projected separation 3,770~au at 398~pc, distances for HOPS~12 and 124 from][]{2020ApJ...890..130T}  from the flat-spectrum protostar HOPS~125 \citep{2016ApJS..224....5F}. Spitzer resolves a scattered light nebula associated with HOPS~124 that is also apparent in high angular resolution HST/NICMOS 1.60 and 2.05~$\mu$m images \citep{2021ApJ...911..153H}. These two bursts were not identified by \citet{2019ApJ...872..183F} or \citet{2021ApJ...920..132P} since HOPS~12 and 124 are blended with their companions in the ALLWISE point source catalog. Furthermore,  HOPS~124 was in the middle of its rise during the 2010 WISE observations (see Sec.~\ref{sec:lc}).  


\subsection{The 3--5~$\mu$m Light Curves between 2004 and 2020}
\label{sec:lc}

\begin{figure*}[t!]
\includegraphics[width=18cm]{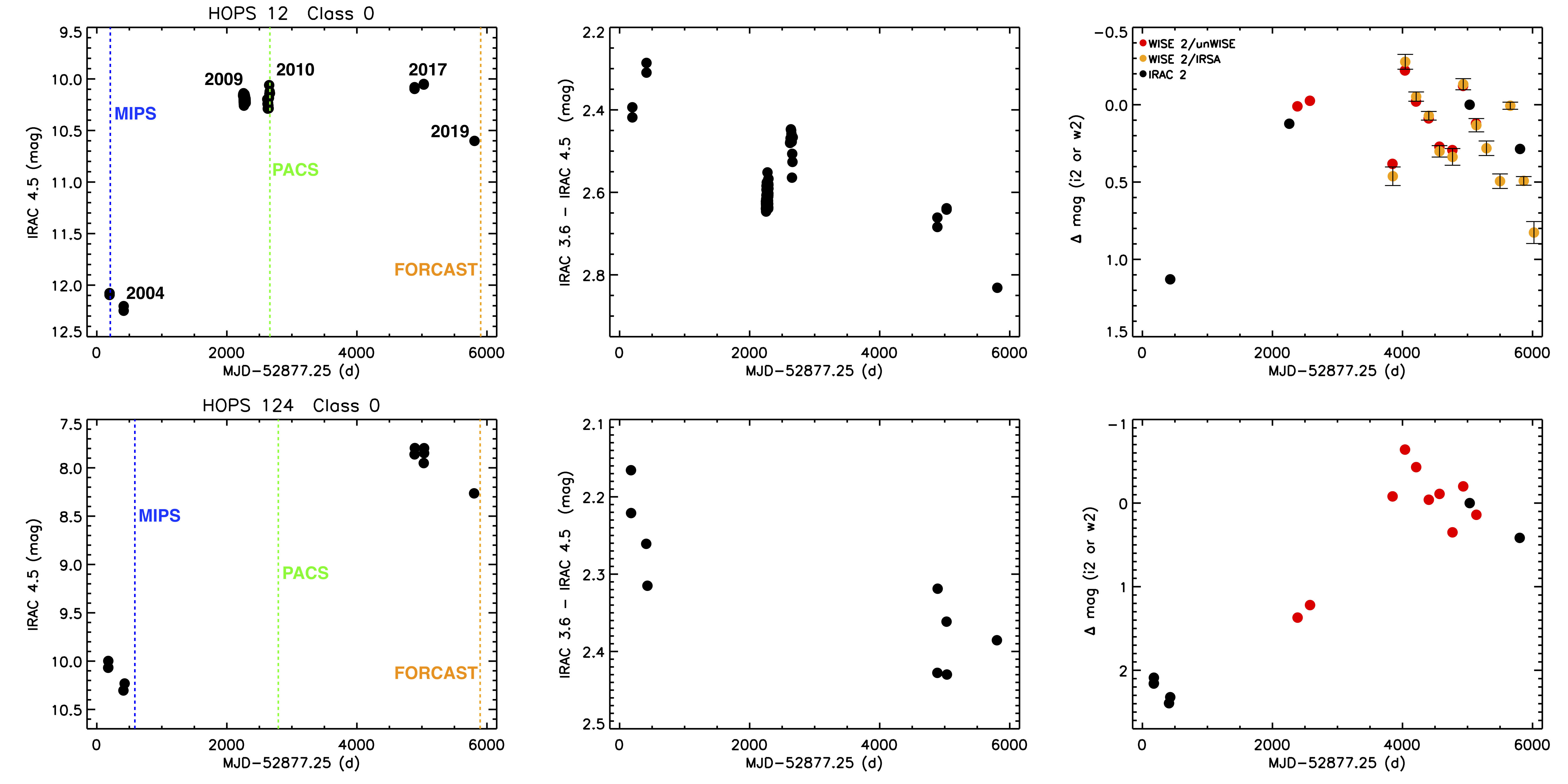}
\caption{Left:\ 4.5 \micron\ light curves of HOPS 12 (top) and HOPS 124 (bottom). Included are the cryo observations from 2004, YSOVAR data from 2009--2010, OrionTFE measurements from 2016--2017, and Spitzer Beyond observations from 2019. HOPS 124 did not fall within the YSOVAR field. Dashed lines show the times of the MIPS, PACS, and FORCAST observations. Middle:\ Spitzer $[3.6]-[4.5]$ colors as a function of time. Right:\ Spitzer (4.5 \micron),  WISE, and NEOWISE (4.6 \micron) light curves; these have been measured relative to the averaged Spitzer 2016--2017 magnitude (for Spitzer) or the averaged magnitude of the NEOWISE data bracketing the averaged date of the OrionTFE data (for WISE/NEOWISE). The Spitzer photometry for the 2004, 2010, and the 2016--2017 epochs have been averaged. The HOPS~12 points are for the combined East and West sources.}
\label{fig:lc}
\end{figure*}

The light curves for the two new outbursts are shown in Figure~\ref{fig:lc}. HOPS~12 brightened between 2004 and 2017 by 1.66 mag at 3.6 $\micron$ and by 2.09 mag at 4.5 $\micron$, corresponding to factors of 4.6 and 6.8 in flux density. The light curve shows that HOPS~12 rose rapidly between 2004 and 2009, sustained its brightness through 2017, and faded by $\sim 0.5$~mag by 2019. Between 2004 and 2017, HOPS~124 brightened by 2.2 mag at 3.6~\micron\ and 2.3 mag at 4.5 \micron, or factors of 7.5 and 8.2. HOPS 124 did not fall within the YSOVAR field and was not observed between 2004 and 2016--2017 by Spitzer. The Spitzer Beyond observations show a $\sim$ 0.5 mag fade between 2017 and 2019. 

The middle panels in Figure~\ref{fig:lc} show the $[3.6]-[4.5]$ colors as a function of time. They show that the colors of both protostars became $\sim 0.5$ mag redder during the bursts. This is the opposite of what one would expect if the brightening were due to a decrease in extinction, and we interpret this as evidence that the bursts are due to rises in the luminosities of the protostars.

The rightmost panels in Figure~\ref{fig:lc} show the relative changes in the Spitzer, WISE and NEOWISE photometry. Relative changes are used to minimize offsets due to the different filter profiles of Spitzer and WISE. The Spitzer data are measured relative to the averaged 2016 and 2017 OrionTFE epochs. The WISE/NEOWISE data are measured relative to the average of the two photometry points bracketing the OrionTFE 2016--2017 averaged date.  For HOPS 12, the combined magnitudes of the East and West sources obtained from the convolved Spitzer images (see Sec.~\ref{sec:spitzer_data}) are displayed since the two objects are not resolved by WISE.


The HOPS~12 NEOWISE data show that by 2015, the source started fluctuating during its burst with peaks in brightness separated by intervals of approximately two years.  For HOPS~124, the data show that the source is rising in 2010 during the WISE observations, and is approximately midway between the 2004 and 2015-2019 mags.  It is not clear if the WISE observations caught a rapidly rising burst, or whether the source had a more gradual increase. The protostar peaked in 2015, and similar to HOPS~12, fluctuated during its burst.  

In summary, both bursts have lasted for 9 yr, although with a potentially slower rise for HOPS~124. The light curves show $> 0.5$~mag variability during the bursts and that both sources may be declining by 2019. 

The 4.5 \micron\ light curve of the previously known outburst for the Class 0 protostar HOPS~383 is shown in Figure~\ref{fig:383}. Since it is located in the ONC, the source was covered by the YSOVAR data. Its light curve encompasses the entire burst event, which had a duration of $\sim 15$~yr. The observed decline is consistent with the report by \citet{2020A&A...638L...4G}, based on NEOWISE data, that the outburst of HOPS 383 has ended. This protostar shows the largest amplitude change of our Class 0 protostar sample; the change of 3.6 mag at 4.5~$\mu$m corresponds to a factor of 22 change in flux density. 

\subsection{Measuring the Burst Amplitude}

Observations at 19-100~$\mu$m with MIPS, PACS, FORCAST, and  HAWC+ measure changes in the luminosities of the outbursting protostars.  While the 4.5~$\mu$m data trace scattered light from the disks of the protostars, the 19-100~$\mu$m data cover the reprocessed luminosities of the protostars \citep{2003ApJ...598.1079W}.  The fluxes at these wavelengths are more tightly correlated with the luminosities of protostars, with the correlation improving with increasing wavelength \citep{2008ApJS..179..249D}. The shape of the protostellar SED over these wavelengths changes slowly with increasing luminosity, with the peak of the SED trending to shorter wavelengths \citep{1993ApJ...414..676K,2016ApJS..224....5F}. Hence, to reasonable approximation, the fluxes in this wavelength range scale linearly  with changes in luminosity, particulary those in the 70-100~$\mu$m range. 

\begin{figure*}[t!]
\includegraphics[width=\linewidth]{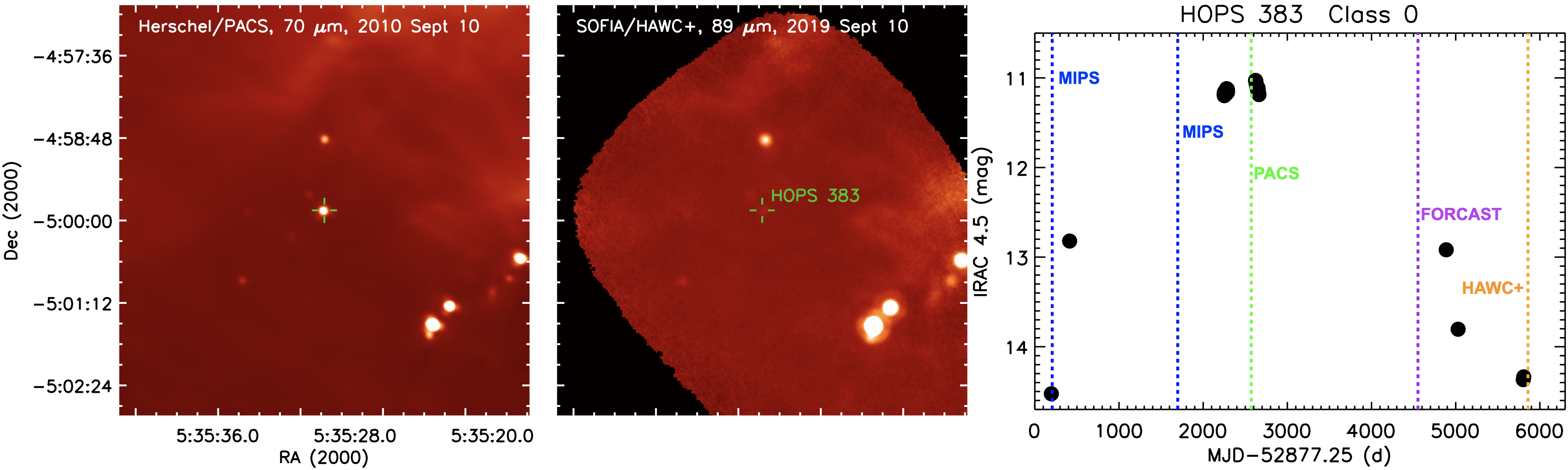}
\caption{Comparison of HOPS 383 and nearby protostars in the 70 \micron\ Herschel/PACS image obtained on 2010 September 10 (left) and the 89 \micron\  SOFIA/HAWC+ image obtained on 2019 September 10 (middle). The location of HOPS 383 is marked ($\alpha=83.87421^\circ$, $\delta=-4.99754^\circ$ J2000).  Right: Light curve of HOPS 383 at 4.5 \micron. The times of the MIPS, FORCAST, PACS and HAWC$+$ observations are shown by the vertical lines.}
\label{fig:383}
\end{figure*}

The 19-100~$\mu$m measurements taken before or after bursts can be compared to those obtained during bursts to estimate the amplitude of each burst, i.e.\ the multiplicative factor by which the flux changes. This amplitude is approximately equal to the amplitude of the luminosity change. The vertical lines in Figure~\ref{fig:lc} show when the 19-100~$\mu$m measurements were made relative to the bursts tracked by the 4.5~$\mu$m light curves.   

The pre-outburst MIPS 70~$\mu$m photometry of HOPS~12 shows a flux of 5.78~Jy, compared to the PACS flux of 15.56~Jy during the burst.  Our deblending of HOPS~12 East and West sources with the PACS data show that 14\% of the 70~$\mu$m flux during the burst, i.e. 2.2~Jy of the total flux, comes from HOPS~12 East. If we assume that HOPS~12~East did not vary between the two epochs, then the pre- and post-outburst fluxes of HOPS~12~West are 3.6 and 13.4~Jy. These show an increase in the flux by a factor of 3.7 at 70~$\mu$m. 

HOPS~12 was also measured by MIPS at 24~$\mu$m pre-outburst (2004) and with FORCAST during the burst (2019; Table~1). Correcting for the differences in the SOFIA and Spitzer bandpasses (see Appendix~A), we find the flux increased by a factor of 2.5 at 24~$\mu$m. The increase is smaller than that measured by the 70~$\mu$m data; this is likely due to the decline in the burst by 2019 when the FORCAST data were obtained. We therefore use the PACS value of 3.7 as the amplitude of the luminosity change for this source.  

HOPS~124 was also measured by MIPS at 24~$\mu$m pre-outburst (2005) and with FORCAST during the burst (2019; Table~2). Again correcting for the different filter wavelengths (Appendix~A), we find the flux increased by a factor of 5.4 at 24~$\mu$m. We use this factor as the amplitude of the luminosity change; however,  since HOPS~124 was declining in 2019, this is likely an underestimate of the peak burst amplitude.  

For HOPS 383, we compared Herschel/PACS data during the burst in 2010 to SOFIA/HAWC+ data taken after the burst in 2019  \citep[Table~2,][]{2016ApJS..224....5F}. 
Linearly interpolating between the 70 \micron\ and 100 \micron\ PACS measurements gives an 89 \micron\ flux density of $20.9$ Jy. 
Comparing that to the 89 \micron\ HAWC+ data (Figure~\ref{fig:383}), we find that the source faded by a factor of 9.7.  

This amplitude is less than the factor of 35 increase at 24~$\mu$m between 2004 and 2008 reported by \citep{2015ApJ...800L...5S}.  In 2016, FORCAST observed HOPS~383 and found that the flux increase was a factor of 40 higher than 2004 (Appendix~A). The discrepency between the 24-25~$\mu$m amplitudes and 70-89 \micron\ amplitude is not understood. The outburst may not have returned to the 2004 levels during the HAWC+ observations.  Alternatively, the 24~$\mu$m fluxes may have been boosted by changes in the disk/envelope during the outburst.  We adopt the 70-89 \micron\ amplitude of 9.7 since it is more conservative and likely more representative of the luminosity change during the outburst. Nevertheless, the FORCAST observations shows that the outburst was still ongoing in 2016. 

\section{Discussion}

\subsection{The Burst Interval}

Among the 92 Class 0 protostars in the Orion sample \citep{2016ApJS..224....5F}, we detected three outbursts with $\ge 2$~mag increases in the Spitzer 3.6 or 4.5~$\mu$m bands. Only two bursts of this magnitude are detected among the 227 more evolved Class I and flat-spectrum protostars in Orion: HOPS~41 in the ONC and V2775~Ori in Lynds 1641 \citep[see Sec~3, locations in][]{2016ApJS..224....5F}. Therefore, the Orion Class 0 protostars have a higher incidence of outbursts than their older counterparts. We infer an outburst rate of (3 bursts) / (92 protostars) / (13 yr) = $2.5 \times 10^{-3}$ yr$^{-1}$ for Class 0 protostars. This corresponds to a burst interval (interval $=$ 1/rate) of one burst every 400 yr for each protostar.   

\begin{figure*}[t!]
\center{\includegraphics[width=\linewidth]{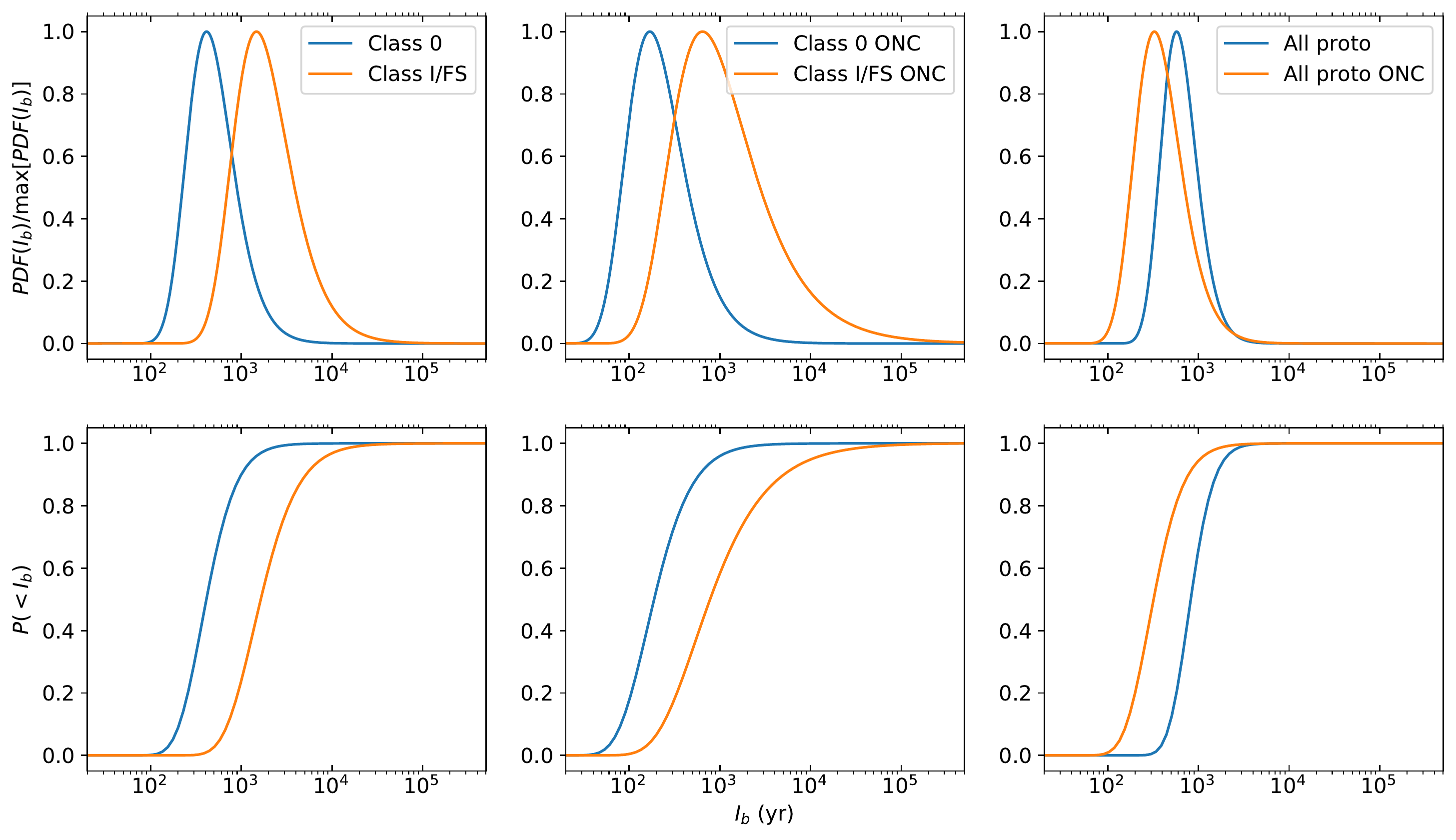}}
\caption{Probability density functions (top) and cummulative probability distributions (bottom) for the burst interval. These are calculated for the entire clouds sample (left:\ 92 Class 0 protostars and 227 Class~I/FS protostars), the YSOVAR field of the ONC (middle:\ 27 Class 0 protostars and 45 Class~I/FS protostars), and all protostars in the clouds and in the ONC (right:\ 319 protostars in the clouds and 72 in the ONC). A 13 year baseline is assumed. These are based on equation~2 from \citet{2019ApJ...872..183F} \citep[also see][]{2015ApJ...808...68H,2019MNRAS.486.4590C}.}
\label{f.rate}
\end{figure*}

From the number of bursts, the number of protostars, and the observed time interval, a probability distribution can be derived for the burst interval \citep{2013MNRAS.430.2910S,2019MNRAS.486.4590C,2015ApJ...808...68H}. Following the approach of \citet{2019ApJ...872..183F}, we present the posterior probability density functions and cumulative probability distributions for the burst intervals in Figure~\ref{f.rate}. These values assume a logarithmic prior; the details of our approach and the dependence of the results on the adopted prior are discussed in Appendix~B.

We calculate burst intervals for three pairs of samples. The first pair is (i) all the Class 0 protostars and (ii) all the more evolved Class I/FS protostars. Since the HOPS 383 outburst subsided by the time of the OrionTFE survey, however, it was only detected since it fell within the YSOVAR observations of the ONC. Thus, protostars outside the ONC that burst after 2004 and declined before 2016 would be missed in our survey. For this reason, the second pair is (i)\ the Class 0 protostars in the ONC and (ii)\ Class~I/FS prototstars in the ONC. This pair has more complete time coverage, but contains a smaller sample of protostars than the first pair. The final pair is (i) all protostars in the entire field and (ii)\ all protostars in the ONC. 

\begin{deluxetable}{lrr}
\tablecaption{Burst Intervals}
\tablewidth{0pt}
\tablecolumns{3}
\tablehead{\colhead{Sample} & \colhead{Median (yr)} & \colhead{95\% CI (yr)\tablenotemark{a}}}
\startdata
Class 0 & 438 & 161 -- 1884 \\
Class I/FS & 1741 & 527 -- 12011 \\
Class 0 in ONC & 199 & 60 -- 1393 \\
Class I/FS in ONC\tablenotemark{b} & 823 & 155 -- 21694 \\
All protostars\tablenotemark{c} & 879 & 400 -- 2515 \\
All protostars in ONC\tablenotemark{c} & 341 & 126 -- 1468 \\
\enddata
\tablenotetext{a}{95\% confidence interval using logarithmic prior}
\tablenotetext{b}{These values are strongly dependent on the prior due to the detection of only one burst.}
\tablenotetext{c}{Class~0, Class~I and Flat~Spectrum combined}
\label{table:burst}
\end{deluxetable}

The median values and 95\% confidence interval from the cumulative distributions are given in Table~\ref{table:burst}. Our burst interval for all protostars is consistent with previous estimates of 1000~yr. The previous estimates used a combination of Spitzer and WISE photometry \citep{2019ApJ...872..183F} and, more recently, NEOWISE photometry \citep{2021ApJ...920..132P} to infer a burst interval. Although the burst intervals from all three studies are consistent, each used different criteria, and the previous studies used lower magnitude thresholds for bursts than our 2~mag cut. Furthermore, the previous studies relied on WISE/NEOWISE data which, primarily due to the lower angular resolution of WISE, missed the HOPS~12 and HOPS~124 bursts. The median burst interval in Table~\ref{table:burst} ($P(t < I_b) = 0.5$) shrinks to 341~yr for the ONC sample alone due to the better time sampling with the inclusion of the YSOVAR data.  Thus, the 1000~yr burst interval may be an overestimate, although due to the small size of the ONC sample, it still falls within the 95\% confidence interval. 

These data also show that Class 0 protostars have a median burst interval of 438~yr, a factor of four shorter than that for the more evolved Class I/FS protostars.  The interval for Class 0 protostars shrinks to 199~yr for the  ONC sample, still a factor of four shorter than that of more evolved protostars in the ONC. To test the significance of these differences, we calculate the p-values assuming $t_b = 879$~yrs using Equation~\ref{eqn:burst_rate}.  For the Class 0 protostars we calculate the probability that the number of bursts is greater than or equal to the number of observed bursts; these are $P(n_{burst} \ge 3) = 15\%$  and $P(n_{burst} \ge 2) = 6\%$ for the clouds and ONC sample, respectively. For the Class I/FS protostars we calculate the values for the number of bursts less than or equal to the observed numbers, these are  $P(n_{burst} \le 2) = 35\%$ and $P(n_{burst} \le 1) = 86\%$. If the burst interval is 342~yrs, as suggested for all protostars using the ONC sample, then $P(n_{burst} \ge 2) = 27\%$ and $P(n_{burst} \le 1) = 49\%$ for the ONC. Thus, we cannot rule out that protostars of all classes have the same burst interval.  

There are caveats to this analysis. Due to degeneracies between inclination, foreground extinction, and evolution, the division into samples based on SEDs may place a few evolved protostars into the Class~0 sample and vice versa. Based on an analysis of models presented in \citet{2013ApJ...767...36S} and \citet{2016ApJS..224....5F}, \cite{2015A&A...577L...6S} found that extinction introduces contamination at less than the 10\% level. It is also possible that some Class~0 protostars are more evolved protostars seen at high inclinations. Only 4 of 75 Class 0 protostars observed in HST 1.6~$\mu$m observations, however, show edge-on inclinations \citep{2021ApJ...911..153H}, and none of the outburst sources appear edge-on. Thus, contamination of the Class~0 sample should have a smaller effect than the statistical uncertainties on the burst interval. A more important caveat is the assumption that all protostars have similar burst rates. This requires that bursts do not result from multiple, distinct mechanisms and that the properties of the protostar, such as binarity, do not affect outbursts \citep{2004ApJ...608L..65R,2013MNRAS.430.2910S,2021MNRAS.507.6061R}. More sophisticated treatments will require larger samples and  spectroscopic data. 


Our measurement of the burst interval for Class~0 protostars can be compared to indirect estimates from the literature. \citet{2019ApJ...884..149H} argued that  N$_2$H$^{+}$ and HCO$^+$ molecules are depleted in the inner envelopes of protostars during bursts as the CO and H$_2$O snowlines move outward, and then reform after the bursts as the snowlines recede. They predict a much longer burst interval of 2400~yr from the fraction of Perseus Class~0 protostars with depletion. Our burst interval is similar to the $310 \pm 150$~yr interval implied by the spacing between ejection events in the outflow from the Class~0 protostar CARMA-7, located in the Serpens South Cluster \citep{2015Natur.527...70P}.  It is also shorter than the 1-3 kyr interval between ejection events for outflows driven by protostars and pre-main sequence stars in nearby clouds \citep{2012MNRAS.425.1380I,2016MNRAS.462.1444F}; this is consistent with a longer interval for more evolved sources. These data are evidence for a connection between the outbursts and ejection events in outflows. 

\subsection{The Fraction of Mass Accreted During Bursts}

A primary goal of studies of episodic accretion is to constrain the fraction of the mass accretion that occurs during bursts ($M_b/M$). We can estimate that ratio for the Class 0 phase using the  interval between bursts ($I_b$), the  duration of bursts ($d_b$), and the  ratio of the accretion rate during bursts to the accretion rate during quiescent intervals between bursts ($A$). From \citet{2019ApJ...872..183F}, 

\begin{equation}
M_b/M=\frac{A d_b / I_b}{(A-1) d_b / I_b + 1}.
\end{equation}

\noindent
We calculate the fraction for the Class 0 protostars using the rate of bursts determined above, $I_b = 438$~yr and a burst duration of $d_b = 15$ yr.  For $A$, we adopt burst amplitudes of 3.7 for HOPS~12, 5.4 for HOPS~124, and 9.7 for HOPS~383; this assumes that all the luminosity is coming from accretion and the accretion luminosity scales with the bolometric luminosity. The resulting $M_b/M$ values are 12, 16 and 26\%. If we reduce $I_b$ to the ONC value of 199 yr, the $M_b/M$ values are 23, 31 and 44\%. If we increase $I_b$ to 1884 yr, the upper confidence interval value for the full sample, then $M_b/M$ are 3, 4 and 7\%.

Since the intrinsic luminosities of the protostars also contribute to the observed luminosities, the ratio of pre-outburst to post-outburst luminosities is a lower bound on $A$. If the entire pre-outburst luminosity of HOPS~383, 0.5~L$_{\odot}$, is due to its intrinsic luminosity, then  $M_b/M =$ 100\%. From the $L_{bol}$ values in \citet{2016ApJS..224....5F} and the burst amplitudes, we estimate the pre-outburst luminosities of HOPS~12 and HOPS~124 are approximately 2 and 9.5~L$_{\odot}$. If their intrinsic luminosities contribute 1~L$_{\odot}$ to their total luminosities, then $M_b/M$ may be as high as 39\% and 33\%, respectively, assuming $I_b$ is 199~yr.   Thus,  $M_b/M$ may be as low as 3--39\% for HOPS~12, or as high as 7--100\% for HOPS~383.

These data show bursts can play a significant role in stellar mass assembly, particularly in the Class~0 phase. Variability with lower amplitudes than those considered here may also play a role. The JCMT transient survey measured the contribution of lower amplitude accretion events over a 4~yr interval  \citep{2021ApJ...920..119L}. These events contributed about 12\% percent of the total mass accreted onto Class~0 protostars. (Note that \cite{2021ApJ...920..119L} define the fraction of mass accreted during bursts as the excess accretion over the quiescent level divided by the total mass accreted. Since we define the fraction of mass accreted as all  the mass accreted during bursts divided by the total accreted mass, we have revised their estimate to be consistent with our definition.) We will explore the contribution of lower amplitude bursts in a subsequent paper (W.~Zakri et al., in prep.).

\subsection{Implications for Burst Mechanisms}

Outbursts are thought to result from the steady accumulation of mass in disks when the rate of infall from the protostellar envelopes onto the disks exceeds the rate of accretion from the disks onto the protostars. As the disks grow more massive, gravitational and/or thermal instabilities lead to episodes of rapid accretion onto the central protostar \citep{1994ApJ...427..987B,1995ApJ...444..376B, 2015ApJ...805..115V, 2009ApJ...701..620Z, 2014ApJ...795...61B}. 

The high rate of bursts of Class~0 protostars supports this picture since the rate of mass infall onto the disks is highest during the Class 0 phase \citep{2016ApJS..224....5F,2017ApJ...840...69F}. Furthermore, VLA and ALMA observations by \citet{2020ApJ...890..130T} find that  disks around Class 0 protostars are systematically more massive than those of Class I/FS protostars, hinting that they are more likely to become gravitationally unstable. Indeed, \citet{2020ApJ...890..130T} find a Toomre-Q value for HOPS~124  indicative of an unstable disk. \citet{2020ApJ...904...78S} also finds the HOPS 383 disk is close to the stability limit; this was based on VLA observations made in 2016 just before the outburst started subsiding. 

Close binary companions have also been proposed as triggers of outbursts \citep{2004ApJ...608L..65R,2013Natur.493..378M,2021MNRAS.507.6061R}; although this mechanism may still require the replenishment of the disk by infall \citep{2016ApJ...830...29G}. So far there is no indication of triggering by binaries. Only one of the  outbursting Class 0 protostars, HOPS~124, show potential evidence for a close companion; in this case, substructure in its disk resolved by ALMA \citep{2020ApJ...902..141S}. Furthermore, since Class 0 protostars do not show a higher incidence of close binaries, triggering by binaries cannot explain a higher rate of outbursts in the Class~0 phase \citep{2021arXiv211105801T}.



\section{Conclusions}

\noindent
1. We compared Spitzer mid-IR photometry from 2004 to 2017 for 92 Class 0 Orion protostars in Orion and detected 3 bursts of $\ge 2$ mag in the 3.6 or 4.5 \micron\ bands. The bursts of the prototstars HOPS~12 and HOPS~124 are new detections. These join the known outbursts from the Class 0 protostar HOPS 383, and the Class~I/FS protostars HOPS 41 and V2775~Ori. 

\noindent
2. The median probability time interval between bursts is 199-438 yr for the Class 0 protostars; this interval is a factor of 4 less than that estimated for Class I/FS protostars. We cannot rule out, however, that all protostars have a burst interval of 860~yr, as inferred for the combined sample of 5 protostars.  Thus, we find suggestive but not conclusive evidence that the burst interval is smaller for Class 0 than for Class I/FS protostars.  This high rate of bursts can be driven by the rapid infall onto to the disks of the Class 0 protostars. 

\noindent
3. The combined Spitzer, WISE, and NEOWISE photometry show that the bursts last at least 9~yr.  The entire HOPS 383 burst lasted 15~yr, the remaining 2 Class~0 bursts did not end in 2019. There is significant variability during the bursts.  

\noindent 
4. With SOFIA, Spitzer, and Herschel data from 19 to 89 \micron, we estimate burst amplitudes for the three protostars. Together with the burst interval and duration, these suggest a significant fraction of the mass during the Class 0 phase is accreted during bursts. The fractions vary from source to source, and may be as low as 3\% for HOPS~12 and as high as 100\% for HOPS~383. While bursts are not required to resolve the protostellar luminosity problem, they can play a significant role in mass accretion onto low-mass protostars.

\acknowledgments

We thank the referee for their  insightful comments. 
This work uses observations from the \textit{Spitzer Space Telescope}, operated by JPL/Caltech under a contract with NASA. This paper also uses data from the \textit{Wide-field Infrared Survey Explorer}, a joint project of the University of California, Los Angeles, and JPL/Caltech, funded by NASA.  Observations were also made with the NASA/DLR Stratospheric Observatory for Infrared Astronomy (SOFIA). SOFIA is jointly operated by the Universities Space Research Association, Inc.\ (USRA), under NASA contract NNA17BF53C, and the Deutsches SOFIA Institut (DSI) under DLR contract 50 OK 0901 to the University of Stuttgart. Herschel is an ESA space observatory with science instruments provided by European-led Principal Investigator consortia and with important participation from NASA. Finally, this work makes use of the NASA/IPAC Infrared Science Archive, operated by JPL/Caltech under a contract with NASA. STM and RAG were supported by the NASA ADAP grant 80NSSC19K0591, and STM was supported by the NASA ADAP grant 80NSSC20K0454. RP was supported by the NASA ADAP grant 80NSSC18K1564. Support for WJF was provided by NASA through award \#07\_0200 issued by USRA. AS gratefully acknowledges funding support through Fondecyt Regular (project code 1180350), from the ANID BASAL project FB210003, and from the Chilean Centro de Excelencia en Astrof\'isica y Tecnolog\'ias Afines (CATA) BASAL grant AFB-170002. M.O. acknowledges support from the Spanish MINECO/AEI AYA2017-84390-C2-1-R (co-funded by FEDER) and MCIN/AEI/10.13039/501100011033 through the PID2020-114461GB-I00 grant, and from the State Agency for Research of the Spanish MCIU through the ``Center of Excellence Severo Ochoa'' award for the Instituto de Astrof{\'i}sica de Andaluc{\'i}a (SEV-2017-0709). This work was completed while STM was a Fulbright Scholar hosted by AS at the Universidad de Concepc\'ion.  The National Radio Astronomy Observatory is a
facility of the National Science Foundation operated
under cooperative agreement by Associated Universities, Inc.

\appendix

\section{Color Corrections}

\begin{figure}
\centering
\includegraphics{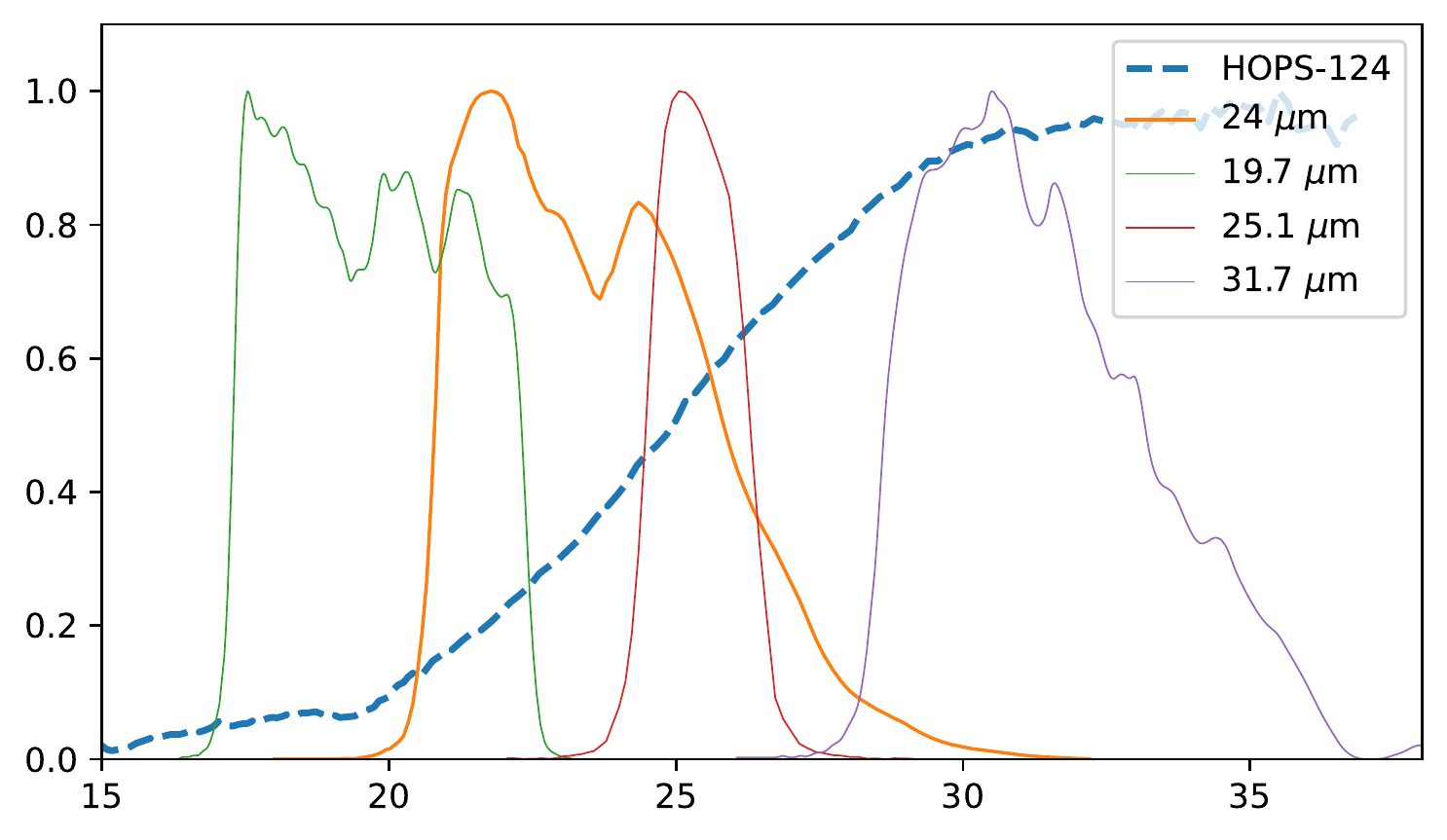}
\caption{The transmission profiles for the FORCAST 19.7~$\mu$m filter, FORCAST 25.3~$\mu$m filter, FORCAST 31.7~$\mu$m filter and MIPS 24~$\mu$m filter. The IRS spectrum of HOPS 124 is also displayed.  All have been normalized so that the peak value is unity.}
\label{fig:filter}
\end{figure}

To compare the FORCAST 19.7~$\mu$m, 25.3~$\mu$m and 31.7~$\mu$m photometry to the MIPS 24~$\mu$m photometry, it is essential to correct for the different wavelength ranges of the three bandpasses, all of which overlap with the MIPS bandpass (Figure~\ref{fig:filter}). This must take into account the spectral shape of the observed protostars, the adopted spectral shapes for the calibration of the FORCAST and MIPS photometry, and the nominal wavelengths adopted for each filter. We estimate the correction by assuming that the burst does not change the spectral shape of the protostar. We use the IRS spectra of HOPS~12 and HOPS~124    \citep[Figure~\ref{fig:filter};][]{2016ApJS..224....5F}. For HOPS~383, which does not have an IRS spectrum, we use the best fit radiative transfer model of the SED from \citet{2016ApJS..224....5F}. We start by determining the flux density in each of the bandpasses based upon the IRS spectrum. In the FORCAST photometry, the calibration of the delivered data assumes a flat-spectrum source. Since the units of flux density for the calibrated FORCAST photometry and the IRS spectrum are in Janskys (i.e., they are in $F_{\nu}$ instead of $F_{\lambda}$), yet the IRS spectra are parameterized by wavelength, we use the following equation for the FORCAST flux density 

\begin{equation}
    F^{\rm IRS/FOR}_{\nu}(\lambda_0) = \frac{\int (1/\lambda) F^{\rm IRS}_{\nu}(\lambda) T(\lambda) d\lambda}{ \int (1/\lambda_0)  T(\lambda) d\lambda},
\end{equation}

\noindent
where $F^{\rm IRS/FOR}_{\nu}(\lambda_0)$ is the inferred flux density for the FORCAST filter, $F^{\rm IRS}_{\nu}(\lambda)$ is the flux density from the IRS spectra,  $T(\lambda)$ is the transmission in electrons per photon, and $\lambda_0$ is the adopted nominal wavelength for the filter (i.e., 19.7, 25.3 or 31.7~$\mu$m). The MIPS calibration assumes a source with a Rayleigh-Jeans spectrum \citep{2007PASP..119..994E}. The resulting flux density inferred for MIPS at 23.675~$\mu$m (the nominal wavelength) is

\begin{equation}
    F^{\rm IRS/MIPS}_{\nu}(\lambda_0) = \frac{\int (1/\lambda^2) F^{\rm IRS}_{\nu}(\lambda) R(\lambda) d\lambda}{ \int  (\lambda_0^2/\lambda^4) R(\lambda) d\lambda},
\end{equation}
\noindent
where $F^{\rm IRS/MIPS}_{\nu}$ is the flux density from the MIPS observations and $R(\lambda)$ is the response in electrons per energy.  

Assuming the shape given by the IRS spectrum remains constant, then the ratio of the two flux densities can be used to estimate the contemporaneous MIPS flux densities from the actual FORCAST photometry. The estimated MIPS flux density is given by

\begin{equation}
F^{\rm FOR/MIPS}_{\nu}(23.675~\mu m) = \frac{F^{\rm IRS/MIPS}_{\nu}(23.675~\mu m)}{F^{\rm IRS/FOR}_{\nu}(25.3~\mu m)} F^{\rm FOR}_{\nu}(25.3~\mu m),
\end{equation}

\noindent
where $F^{\rm FOR}_{\nu}(25.3~\mu m)$ is the observed flux density. The burst amplitude is then given by

\begin{equation}
    A = \frac{F^{\rm FOR/MIPS}_{\nu}(23.675~\mu m)}{F^{\rm MIPS}_{\nu}(23.675~\mu m)}.
\end{equation}

\noindent
Similar equations can be used with the observed flux densities in the two other FORCAST filters. 

Using the flux densities in Table~\ref{t.long_phot} for the FORCAST and MIPS data, we estimate the burst amplitudes in Table~\ref{t.cc}. While the amplitudes determined from the  25.3~$\mu$m and  31.7~$\mu$m  photometry are consistent, the amplitudes derived from the 19.7~$\mu$m filter are higher. This is likely due to the breakdown of our assumption that the pre-outburst spectrum approximates the spectrum during the burst at this wavelength.  Given that the 25.3~$\mu$m band is closest to the the nominal wavelength of the MIPS 24~$\mu$m filter, we adopt these values in our analysis of the bursts.

\begin{deluxetable}{cccc}
\tablecaption{FORCAST Burst Amplitudes\label{t.cc}}
\tablewidth{0pt}
\tablecolumns{4}
\tablehead{\colhead{Filter} & \colhead{19.7~$\mu$m} &  \colhead{25.3~$\mu$m} & \colhead{31.7~$\mu$m}}
\startdata
HOPS~12 & 3.8 &  2.5 & 2.5 \\
HOPS~124 & 8.5 & 5.4 & 5.1 \\
HOPS~383 &  & 40 &  \\
\enddata
\label{table:amplitude}
\end{deluxetable}


\section{Model and Posterior Probability Distribution}

\begin{deluxetable}{lccc}[b]
\tablecaption{Burst Intervals as a Function of Prior}
\tablewidth{0pt}
\tablecolumns{3}
\tablehead{\colhead{Sample} & \colhead{Uniform (yrs)\tablenotemark{a}} & \colhead{Logarithmic (yrs)\tablenotemark{a}} & \colhead{Jeffries (yrs)\tablenotemark{a}}}
\startdata
Class 0 & 701 (248 - 3310) & 438 (161-1884) & 371 (167 - 1086) \\
Class I/FS & 4203 (979 - 51391) & 1741 (527-12011) & 1350 (531 - 5130) \\
Class 0 in ONC & 487 (113 - 6502) & 199 (60-1393) & 155 (61 - 590) \\
Class I/FS in ONC & 22128 (721 - 366828) & 823 (155-21694) &  485 (148 - 3287)\\
All protostars & 1120 (531 - 3011) & 879 (400-2515) & 796 (418 - 1799) \\
All protostars in ONC & 546 (193 - 2579) & 341 (126-1468) & 289 (130  - 846)\\
\enddata
\tablenotetext{a}{median value and 95\% confidence limits in parentheses}
\label{table:burst_prior}
\end{deluxetable}

\citet{2019ApJ...872..183F} presented a model for the number of outbursts where the  probability of the number of bursts, $n_b$ is given by

\begin{equation}
    P(n_{b}| I_b,\Delta T,N_{p}) = \frac{N_p!}{n_{b}! (N_{p}-n_{b})!} (1-e^{-\Delta T/I_b})^{n_{b}} (e^{-\Delta T/I_B})^{(N_{p}-n_{b})}.
    \label{eqn:burst_rate}
\end{equation}

\noindent 
In this model, $I_b$ is the interval between bursts ($I_b = 1/R_b$ where $R_b$ is the rate of bursts), $N_P$ is the number of protostars, and $\Delta T$ is the time interval over which bursts were monitored.  The probability of detecting one or more bursts for a given protostar is $(1-e^{-\Delta T/I_b})$; this value goes to one as $\Delta T/I_b$ goes to infinity. In the limit $\Delta T/I_b << 1$, this model can be approximated by a binomial distribution \citep{2015ApJ...808...68H} or Poisson distribution \citep{2019MNRAS.486.4590C}. We have determined the posterior probability distribution for $P(n_b)$ for three different uninformative priors: a uniform prior, a logarithimic prior, $P(I_b) \propto 1/I_b$, and a Jeffries prior, $P(I_b) \propto 1/I_b^{1.5}$ when $\Delta T/I_b << 1$. In all cases, the prior probability drops to 0 when $I_b > 0.5$~Myr, the lifetime of a protostar.

In Table~\ref{table:burst_prior}, we compare the median value and 95\% confidence intervals obtained from the 2.5\% and 97.5\% points of the cumulative distributions.  The uniform weighting increases the median value of $I_b$; this prior increases by a factor of ten for each successive log$_{10}$ interval of $I_b$, and hence it is weighted to longer values of $I_B$. In contrast, the Jeffries prior gives the shortest values of $I_b$.  Since it is appropriate  to use a logarithmic prior for determining the order of magnitude of the burst interval, we adopt that prior in the paper.  We find that the logarithmic and Jeffries priors produce similar posterior probabilty distributions. Furthermore, the median $I_b$ values for the uniform prior are within the logarithmic prior's confidence intervals. Only for the Class I/FS burst interval in the ONC, which is dependent on one detected burst, is the posterior probability distribution  driven by the prior. In this case, $I_b$ is poorly constrained.

\bibliography{outburst}


\end{document}